\documentclass{emulateapj}

\slugcomment{Draft Version Jun. 17, 2008: Accepted for Publication in ApJ}

\usepackage{times}
\usepackage{bm}

\newcommand{\bolB}{{\bm  B}}

\shorttitle{A Numerical Model of Hercules A by Magnetic Tower}
\shortauthors{Nakamura et al.}

\begin{document}

\title{A Numerical Model of Hercules A by Magnetic Tower: \\ 
Jet/Lobe Transition, Wiggling, and the Magnetic Field Distribution}

\author{Masanori Nakamura\altaffilmark{1}, Ian L. Tregillis \altaffilmark{2}, Hui Li\altaffilmark{1},
and Shengtai Li\altaffilmark{3}}
  \altaffiltext{1}{Theoretical Astrophysics, MS B227, Los Alamos
  National Laboratory, NM 87545; nakamura@lanl.gov}
  \altaffiltext{2}{Applied Physics, MS F699, 
  Los Alamos National Laboratory, NM 87545}
  \altaffiltext{3}{Mathematical Modeling and Analysis, MS B284, 
  Los Alamos National Laboratory, NM 87545}

\begin{abstract}
We apply magnetohydrodynamic (MHD) modeling to the radio galaxy Hercules
A for investigating the jet-driven shock, jet/lobe transition, wiggling,
and magnetic field distribution  associated with this source.  The model
consists of magnetic  tower jets in a galaxy  cluster environment, which
has been discussed in a series of our papers.  The profile of underlying
ambient gas plays an important role in jet-lobe morphology.  The balance
between the magnetic pressure generated by axial current and the ambient
gas pressure  can determine the lobe  radius.  The jet  body is confined
jointly by  the external  pressure and gravity  inside the  cluster core
radius $R_{\rm  c}$, while  outside $R_{\rm c}$  it expands  radially to
form  fat  lobes  in  a  steeply  decreasing  ambient  thermal  pressure
gradient.  The  current-carrying jets  are responsible for  generating a
strong,   tightly   wound  helical   magnetic   field.   This   magnetic
configuration will be unstable  against the current-driven kink mode and
it visibly grows  beyond $R_{\rm c}$ where a  separation between the jet
forward  and return  currents  occurs.  The  reversed  pinch profile  of
global  magnetic  field  associated  with  the jet  and  lobes  produces
projected $\bolB$-vector distributions aligned with the jet flow and the
lobe edge. AGN-driven shock powered  by the expanding magnetic tower jet
surrounds the jet/lobe  structure and heats the ambient  ICM.  The lobes
expand subsonically; no  obvious hot spots are produced  at the heads of
lobes.  Several  key features in  our MHD modeling may  be qualitatively
supported by the observations of Hercules A.
\end{abstract}

\keywords{galaxies:individual: Hercules A --- galaxies: active  
---  galaxies: jets  --- methods:  numerical   ---  MHD}  

\section{INTRODUCTION}  
\label{sec:INT} In the present paper,  we continue our discussion of the
dynamics  of  extragalactic  jets  in cluster  environments  within  the
framework of the ``magnetic  tower'' model \citep[]{LB94, L96}, which we
have  analyzed  in  a   series  of  papers.   Magnetohydrodynamic  (MHD)
mechanisms are  frequently invoked to model  the launching, acceleration
and  collimation  of  astrophysical  jets \citep[see,  {\it  e.g.},][and
references  therein]{F98,  M01}.   An underlying  large-scale  (coronal)
poloidal  field for  producing the  magnetically driven  jets  is almost
universally assumed in  many theoretical/numerical models.  However, the
origin and existence of such  a galactic magnetic field are still poorly
understood.

In   contrast   with   the   large-scale   field   models,   Lynden-Bell
\citep[]{LB94,  L96}  examined the  expansion  of  the local  force-free
magnetic loops anchored to the star  and the accretion disk by using the
semi-analytic  approach.   Twisted  magnetic  fluxes  due  to  the  disk
rotation make the magnetic loops  unstable and splay out at a semi-angle
60\degr \  from the rotational  axis of the disk.   Global magnetostatic
solutions of  magnetic towers with  external thermal pressure  were also
computed  by   \citet[]{L01}  using  the   Grad-Shafranov  equation  in
axisymmetry  \citep[see also,][]{L02, LR03,  UM06}.  Full  MHD numerical
simulations  of magnetic  towers  have been  performed in  two-dimension
(axisymmetric)  \citep[]{R98, T99, U00,  K02, K04a}  and three-dimension
\citep[]{K04b}.   Magnetic towers  are also  observed in  laboratory
experiments \citep[]{HB02, L05}.

The first in our series, \citet[][hereafter Paper I]{L06}, described the
basic assumptions  and approaches in the numerical  modeling of magnetic
tower  jets.    The  evolution   of  the  tower   jets  in   a  constant
density/pressure  background  was examined  there.   The  second in  our
series,  \citet[][hereafter  Paper   II]{N06}  investigated  the  global
structure  of  magnetic   tower  jets  in  a  gravitationally-stratified
atmosphere,  in   terms  of  the   MHD  wave  structure,   radial  force
equilibrium, and  collimation.  A large current flowing  parallel to the
jet  bulk flow  plays  an essential  role  in the  determining the  lobe
radius,  as does  the background  pressure  profile.  The  third in  our
series,  \citet[][hereafter Paper  III]{N07} investigated  the stability
properties  of magnetic  tower jets  on cluster  scales.  Current-driven
instabilities are  responsible for the  non-axisymmetric structures; the
external  ``kink'' ($m=1$)  mode grows  outside $R_{\rm  c}$,  while the
internal  ``double  helix''  ($m=2$)  mode  grows  predominantly  inside
$R_{\rm c}$.  This  is the forth in our series of  papers to examine the
nonlinear magnetic tower jets. Here we examine our model's applicability
to  an individual  source, Hercules  (Herc)  A (3C  348), by  performing
three-dimensional MHD simulations.

Herc A  is one of  the most powerful  double-lobed radio sources  in the
sky.  Its total  energy content is approximately $3  \times 10^{60}$ erg
and it is  identified with the central compact diffuse  (cD) galaxy of a
cluster  \citep[]{SH93} at  a low  redshift of  $z=0.154$ \citep[]{S99}.
The X-ray luminosity of $4.8 \times 10^{44}$ erg s$^{-1}$ \citep[]{GL04}
is  thought to be  due to  bremsstrahlung of  the very  high temperature
intracluster medium (ICM).   Herc A has a peculiar  radio structure; the
morphology looks  like a  classical double-lobed Fanaroff-Riley  type II
(FR II)  radio source  \citep[]{FR74}, but it  has no  compact hotspots,
instead   showing  an   unusual  jet-dominated   morphology  as   in  an
Fanaroff-Riley type I  (FR I) radio source, which  was first revealed by
\citet[]{DF84}.  Furthermore, the two jets in Herc A are quite different
in  appearance.  The  eastern jet  appears to  have  continuous twisting
flows, while  the western  jet leads to  a unique sequence  of ``rings''.
The jets in Herc A are well-collimated initially but they flare suddenly
at a certain distance from the nucleus.  An explanation for the presence
of such different  structures in the same source  is still unresolved in
the literature.  Herc  A is not a typical FR I  source, either; its jets
are well-collimated and has knot components.  Therefore, Herc A might be
classified an intermediate case: FR I/II.

Various models  have been proposed  for explaining the formation  of the
jets and rings  in Herc A, including both  kinematic \citep[]{M88, MS96,
SM02}  and dynamic  \citep[]{M91, S02}  points of  view.   By performing
axisymmetric,  two-dimensional  hydrodynamic simulations,  \citet[]{M91}
proposed a model  in which an {\em over}-dense  and {\em over}-pressured
jet undergoes  a sudden expansion when it  becomes highly over-pressured
compared  to the background  gas at  a certain  distance from  the core.
This causes  a subsequent  conical expansion, giving  rise to  the ``ice
cream cone'' shape of the lobe. However, at present, there is nothing in
the  X-ray  observations  to  suggest  that the  jet  becomes  thermally
over-pressured  as  required  in  this  model.   In  \citet[]{S02},  the
formation of  ring structure  in the western  jet is interpreted  as the
result  of  ribbon-like,  annular  shocks propagating  through  the  jet
cocoon.  Also, these  authors conclude that the absence  of hot spots in
Herc A results  from the dynamics of turbulent  and entraining backflows
in the radio lobe.

Recently,  \citet[]{GL03}  have revealed  detailed  structures of  total
intensity, spectral index, polarization, and projected magnetic field in
Herc A by  using multifrequency VLA imaging.  The  jet disruption in the
eastern  lobe  is clearly  visible,  implying  flow instabilities.   The
projected magnetic field, after correcting for Faraday rotation, closely
follows the edges  of the lobes, the jets, and the  rings; the two lobes
have generally  similar field patterns.  This  observational picture may
disagree  with the  turbulent structure  inside the  lobes  suggested by
\citet[]{S02}.

Magnetic field distributions associated  with the extragalactic jets and
their  comparisons with  the observations  have been  examined  by using
Stokes  parameters, Faraday  rotation  measure (RM),  and the  projected
magnetic  field ($\bolB$-vector)  \citep[e.g.,][]{L81,  C89, C92,  HR99,
HR02,  LCB06}. In  general, helical  magnetic fields  produce asymmetric
transverse  brightness and polarization  profiles; they  are symmetrical
only if  the field is purely toroidal  or the jet is  at 90$^{\circ}$ to
the line of sight  \citep[]{L81}.  \citet[]{LCB06} conclude that, to the
first approximation,  the field  configuration in FR  I jet over  10 kpc
scales is a mixture of toroidal and longitudinal components.

\citet[]{GL04}'s {\it ROSAT} X-ray  observations of the intracluster gas
in the Herc A cluster  have revealed extended X-ray emission coming from
a  compact source  at the  center of  cluster, out  to a  radius  of 2.2
Mpc. The  azimuthally-averaged X-ray surface brightness  profile is well
fitted by a modified King ($\beta$) model \citep[]{K72, CF76}, with core
radius  $R_{\rm c}=121  \pm 10$  kpc and  $\beta=0.74 \pm  0.03$.  Their
X-ray analysis supports the conclusion  that the inner jets are confined
thermally by the cluster core  gas: thermal pressure in the cluster core
is almost 10 times larger than the equipartition pressure in the eastern
jet  with the standard  assumptions (including  no protons).   The radio
lobes  are  largely  positioned  {\em  beyond} the  X-ray  core  radius,
allowing  for  projection, so  they  are  expanding  essentially into  a
power-law atmosphere  with density falling  as $R^{- 3 \beta}  \sim R^{-
2.22}$, quite close to the $R^{\-2}$  profile needed to give the lobes a
self-similar structure \citep[]{F91}.

These observational evidence motivate us to apply our dynamical magnetic
tower  jet model to  Herc A.  To our  knowledge, no  MHD model  has been
applied to Herc A before though there are several hydrodynamic models as
we have  already referred. Radio observations suggest  that the magnetic
fields need to be taken  into account to discuss about jet/lobe dynamics
in Herc  A. We believe that  a key clue to  understanding the transition
between the jet and the lobe may be whether or not the inner jets become
over-pressured  during their  propagation in  the X-ray  thermal cluster
gas.   However, the jet  internal pressure  may be  magnetically, rather
than  thermally, dominated.   This  is in  contrast  to previous  models
\citep[]{M91}.  Of particular interest here are the jet/lobe transition,
wiggling,  and  the  magnetic  field distribution  associated  with  the
eastern jet of Herc A, which  are not yet fully discussed among the past
hydrodynamic models of Herc A. We  do not discuss the formation of rings
in the western jet in this paper.

This paper is  organized as follows.  In \S 2,  we outline our numerical
methods.  In \S  3, we describe our numerical  results.  Discussions and
conclusions are given in \S 4 and \S 5.

\section{NUMERICAL METHODS AND MODEL ASSUMPTIONS}
We  solve the  nonlinear system  of time-dependent  ideal  MHD equations
numerically  in a  3-D Cartesian  coordinate system  $(x,\,y,\,z)$.  The
basic  numerical  treatments (including  the  MHD  numerical scheme)  is
essentially the  same as that in Papers  I - III.  We  assume an initial
hydrostatic  equilibrium  in   the  gravitationally  stratified  medium,
adopting an isothermal  King model.  The magnetic flux  and the mass are
steadily  injected in  a  central  small volume  during  a certain  time
period.   Since the injected  magnetic fields  are not  force-free, they
will  evolve as  a  ``magnetic  tower'' and  interact  with the  ambient
medium.  The dimensionless system of MHD equations is integrated in time
by  using  the  TVD  upwind  scheme  \cite[]{LL03}.   Computations  were
performed  on the  parallel Linux  clusters at  the Los  Alamos National
Laboratory.

\subsection{Numerical Set Up}
We normalize physical quantities with the unit length scale $R_{0}$, the
unit density $\rho_{0}$, and the sound speed $C_{\rm s0}$ as the typical
speed  in   the  system.   Other  quantities  are   derived  from  their
combinations, {\it e.g.}, the  typical time $t_{0}$ is $R_0/C_{\rm s0}$,
etc. In this paper, we  use some observed quantities associated with the
source Herc  A \citep[][hereafter GL04]{GL04} in order  to determine our
normalization units.   A central electron density  $n_{\rm e}=1.0 \times
10^{-2}$ cm$^{-3}$ is adopted in GL04, suggesting quite a dense cluster,
as such densities in clusters are typically of order $10^{-3}$ cm$^{-3}$
\citep[]{JF84}.   This provides a  unit density  $\rho_{0} =  1.7 \times
10^{-26}$ g  cm$^{-3}$.  A single-temperature fit with  $kT=4.25$ keV is
assumed  for  the  $\beta$-model  cluster,  giving  a  unit  temperature
$T_{0}=4.9 \times 10^{7}$ K, a unit pressure $p_{0}=2.3 \times 10^{-10}$
dyn cm$^{-2}$, and a unit  velocity (sound speed) $C_{s0} (\equiv \gamma
p_{0} /  \rho_{0}) =1.2 \times 10^{8}$  cm s$^{-1}$.  Here,  we choose a
unit  length $R_{0}=30$ kpc,  and therefore,  a unit  time $t_{0}  = 2.4
\times 10^{7}$  yr.  The unit magnetic  field $B_{0}$ is  $(4 \pi \rho_0
C_{\rm s0}^2)^{1/2}=53.4$  $\mu$G.  In  the King model  we use  here, we
take the cluster core radius $R_{\rm c}$ to be 120 kpc, corresponding to
the normalized core radius 4.0,  and take the exponential slope $\kappa$
to be $1.1$ ($3\beta/2$  with $\beta=0.74$) by adopting the observations
in GL04.

We  use $\rho_0$ and  $p_{0}$ as  the initial  quantities at  the origin
$(x,\,y,\,z)=(0,\,0,\,0)$ and  thus, the  normalized $\rho$ and  $p$ are
set  to   unity  at  the   origin  in  these  simulations.    The  total
computational  domain is  taken to  be $|x|,\,|y|,\,|z|  \leq  15$.  The
number of grid points in the simulations reported here is $360^3$, where
the  grid  points  are assigned  uniformly  in  the  $x$, $y$,  and  $z$
directions.  The simulation  domain is  from $-450$  to $450$  kpc, with
$\Delta x =\Delta y=\Delta z \sim 0.083$, which correspond to $\sim 2.5$
kpc.

The injections of  magnetic flux, mass, and the  associated energies are
the same as those described in  Paper I.  The ratio between the toroidal
to  poloidal  fluxes  of  the  injected fields  is  characterized  by  a
parameter $\alpha = 25$.  The magnetic field injection rate is described
by $\gamma_b$ and is set to  be $\gamma_b=1$.  The mass is injected at a
rate of $\gamma_\rho  =0.1$ over a central volume  with a characteristic
radius  $r_\rho  = 0.5$.   Magnetic  fluxes  and  mass are  continuously
injected for  $t_{\rm inj} = 3.0$,  after which the  injection is turned
off. These parameters correspond to  a magnetic energy injection rate of
$\sim  2.7 \times  10^{46}$ ergs  s$^{-1}$ and  an injection  time $\sim
72$ Myrs.  The  energy injection rate used here is  a similar range to
that  of other  hydrodynamic simulations  with the  jet power  $2 \times
10^{45-46}$ ergs  s$^{-1}$ \citep[]{S02}.   We use the  outflow boundary
conditions at all outer boundaries. Note that for most of the simulation
duration, the waves and magnetic  fields stay within the simulation box,
and all magnetic fields are self-sustained by their internal currents.

\subsection{Formulation of Synthetic Observation Images}
Based  on  our simulations,  we  make  synthetic  observation images  to
compare  with  the  real   observations  of  Herc  A  \citep[][hereafter
GL03]{GL03}.  The synthetic  radio telescope observations presented here
were  produced  using  a  modified  form of  the  synthetic  observation
technique first  described in \citet{Tregillis01a}.  Here  we provide an
overview  of the  key points  relevant to  the present  investigation; a
complete    description    of    the    technique   is    provided    in
\citet{Tregillis02e}.

Our  method  utilizes  the  vector  magnetic  field  structures  evolved
self-consistently    within    these     MHD    calculations    and    a
relativistic-electron  momentum  distribution,  $f(p)$, chosen  by  fiat
during  postprocessing.   The  function  $f(p)$ is  defined  over  eight
logarithmically-spaced momentum bins.  Each  bin has an associated slope
value, $q$, and within each bin the distribution is a power law ($f \sim
p^{-q}$).   Slope  values may  vary  between  bins,  however, making  it
possible  to  specify non-power  law  distributions,  such  as would  be
appropriate  for  cases  where  radiative  aging  is  significant.   The
specified  distribution functions  are  free to  vary  from location  to
location within  the simulated flows, as  is expected to be  the case in
real radio galaxies.  We note  that although these distributions are not
derived self-consistently within the computed flows, as has been done in
previous  works  \citep{Tregillis01a, Tregillis04a},  imposing  it as  a
postprocessing  step  does  provide  a  measure of  flexibility  in  our
investigations.

In this paper we choose the  sign convention such that $\alpha > 0$, and
the  flux  density $S_{\nu}  \propto  \nu^{-\alpha}$.   For the  present
investigations, which do not  incorporate the effects of radiative aging
and reacceleration, we applied  a spatially-uniform momentum slope value
$q  = 4.5$  throughout  the  simulated source.   This  corresponds to  a
synchrotron spectral index $\alpha = 0.75$, which is consistent with the
observed  range of  spectral indices  within individual  radio galaxies.
For example, \citet{GL03} mapped  the spectral index distribution in
Herc A  between   1.3   and  4.8   GHz,   finding  $0.6   \lesssim
\alpha^{4.8}_{1.3} \lesssim  2.0$, with  the flattest values  within the
western jet  and the  steepest values along  the lobe edges.   They also
found $\alpha^{4.8}_{1.3} \approx 0.75$  in portions of the eastern jet.
Our choice,  $\alpha =  0.75$, is also  consistent with  typical hotspot
spectral index values $\alpha^{5.0}_{1.4}$  in the sample of 3CR sources
studied by \citet{AL87}.

The local  number density of  energetic electrons varied with  the local
magnetic  energy density  (i.e.,  $n_e \propto  B^{2}$).  This  approach
represents a  partitioning relationship  between the field  and particle
energies,   such    as   is   posited    by   minimum-energy   arguments
\citep{Miley80}. However,  the overall scaling was chosen  such that the
energy density of relativistic electrons  was far less than the magnetic
energy  density  (i.e.,  we  assume  our  simulated  object  is  out  of
equipartition, which  is consistent with the lack  of dynamical feedback
from the electrons in these calculations).

Given  information  about  the   local  magnetic  field  and  the  local
distribution of energetic electrons, we compute a synchrotron emissivity
$j_{\nu}$  in  every  zone  of  the computational  grid.   As  given  by
\citet{Jones74}, the emissivity is
\begin{equation}
j_s(\nu) = j_{\alpha 0}{4\pi e^{2} \over c}f(p_{s})p_{s}^{q}\left({\nu_{B_{\bot}}
\over
\nu}\right)^{\alpha}\nu_{B_{\bot}}.
\end{equation}
The spectral  index $\alpha$ is  related to the local  electron momentum
index $q$  via $\alpha =  (q-3)/2$, $\nu_{B_{\bot}}=eB\sin{\Omega}/(2\pi
m_ec)$,  where $\Omega$  projects  the  local field  onto  the sky,  and
$j_{\alpha  0}$ is  an  order-unity dimensionless  constant, defined  in
\citet{Jones74}.   For  a   selected  observing  frequency,  $\nu$,  the
distribution, $f(p_{s})$,  and the index,  $q$, are determined  for each
point  on  the numerical  grid  by  establishing  the relevant  electron
momentum from  the relation $p_{s}  = [2\nu/(3\nu_{B_{\bot}})]^{(1/2)}$,
with $p_s$ in units $m_{e}c$.

Once the  synchrotron emissivity is  known for every numerical  zone, we
compute  surface   brightness  maps  for   optically-thin  emission  via
line-of-sight integrations.  Similarly, we can obtain the Stokes $Q$ and
$U$ parameters for  the synchrotron emission, as well  as the correction
for Faraday  rotation through  the source, making  detailed polarimetric
studies possible.  We write  the resultant data in \textsc{fits} format,
and    analyze   it    using    conventional   observational    packages
(\textsc{miriad} and \textsc{karma} \citep{Gooch95}).

\section{RESULTS}
\label{sec:RES}
Our magnetic tower model has been applied to the Herc A.  In particular,
our  goal  is to  reproduce  large scale  structures  in  the jets:  the
jet/lobe transition, the non-axisymmetric deformation, and the projected
magnetic  field distribution  associated  with the  jets  and lobes,  as
observed in GL03.

\begin{figure} 
\begin{center}
\includegraphics[scale=0.65, angle=-90]{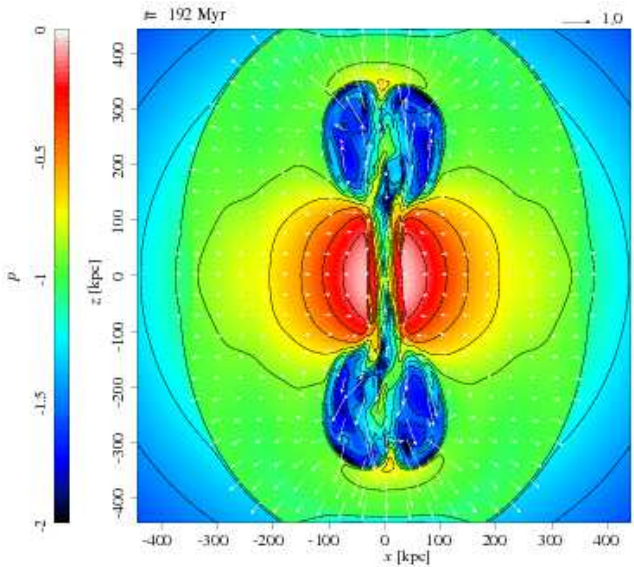} \caption{\label{fig:f1}
Distribution of the gas pressure $p$ (logarithmic scale) with the
velocity field ($V_{x}, \, V_{z}$) ({\em arrows}) in the $x-z$ plane at
192 Myr. The length and time scales are converted into the corresponding
dimensional values.  The cavities (low density and gas pressure) are
formed due to the magnetic tower expansion as corresponding lobes.  The
jets are deformed into the wiggled structures inside the lobes.}
\end{center}
\end{figure}

Figure  \ref{fig:f1} shows a  snapshot of  the gas  pressure $p$  in the
($x-z$)  cross section at  $t=8.0$ (192  Myr).  As  we have  examined in
papers II and  III, the global structure of  the magnetic tower consists
of  a well-collimated ``body''  and radially  extended ``lobes''  in the
gravitationally stratified atmosphere.  A transition from the narrow jet
to the fat  lobe can occur at the cluster  core radius. This interesting
property can  be confirmed in resent observations  (GL03, GL04). Several
key features in these magnetic  tower jets (the MHD wave structures, the
heating  process  at  the  tower  front, the  cylindrical  radial  force
equilibrium at the tower edge,  and the dynamic collimation process) are
similar to  the results of paper  II (see also  Fig.  1 of paper  II for
details of  time evolution). The helically-twisted magnetic  tower has a
closed current  system that includes  a pair of current  circuits.  Each
circuit contains a  forward electric current path (the  jet flow itself,
with its toroidal magnetic field,  toward the lobe) and a return current
(along  some path  back  to the  AGN  core).  The  global  picture of  a
current-carrying  jet   with  a   closed  current  system   linking  the
magnetosphere of the central engine  and the hot spots was introduced by
\citet[]{B78, B06} and  applied to AGN double radio  sources. As seen in
Fig. \ref{fig:f1},  the high-pressure material of the  ambient gas seems
to follow  the edge of the jet  near the bottom of  each lobe.  However,
this is not  the case, but the local  pressure enhancement occurs inside
the lobes.   In the jet  and lobe system,  the axial currents  drive the
current-driven   instabilities;    it   eventually   produces   magnetic
reconnection  process even  in our  ideal MHD  assumption  (the magnetic
field will dissipate  numerically). As a result, some  local heating may
occur,   but  it  never   affects  the   dynamics  of   jet  propagation
dramatically.

For a full 3-D visualization of  the magnetic lines of force, the reader
is referred to  paper III.  The basic behavior  of the helically-twisted
magnetic field  in the  present paper appears  similar to that  in paper
III.   The   magnetic  tower  jet  has  a   well-ordered  helical  field
configuration,  with a tightly-wound  central helix  going up  along the
central axis and a loosely -wound helix coming back at the outer edge of
the  magnetic  tower.  The  outer  edge of  the  magnetic  tower may  be
identified  as  a  tangential  discontinuity without  the  normal  field
component.   The  interior  of   tower  (lobe)  is  separated  from  the
non-magnetized external  gas via this discontinuity. At  the tower edge,
the  outward-directed  magnetic   pressure  gradient  force  is  roughly
balanced with the inwardly-directed  thermal pressure gradient force. On
the  other hand,  at  the core  part  of jet  body,  a quasi-force  free
equilibrium  is  achieved.  In  the  context of  magnetically-controlled
fusion systems, the helical field  in the magnetic tower can be regarded
as the reversed field pinch (RFP) profile.

The  jet  axial  current  and  the ambient  gas  pressure  can  together
determine  the radius  of the  magnetically-dominated lobes  (paper II).
That is, the internal gas pressure plays a minor role in the lobes (even
if  the  thermal  pressure  is  greater than  the  non-thermal  pressure
\citep[]{MO88}   as  is   typically  seen   in  FR   I   radio  galaxies
\citep[]{C03}.  The Alfv\'en speed  becomes large (about three times the
local  sound speed), while  the plasma  $\beta (\equiv  2p/B^2)$ becomes
small  ($\beta \lesssim  0.1$) inside  the density  cavities due  to the
expansion of magnetic fluxes.  As we have already discussed in paper II,
it is  remarkably noted  that the expanding  tower launches  a preceding
hydrodynamic shock in  the ICM, which may be  associated with AGN-driven
shocks seen in recent X-ray observations \citep[]{F05, N05, F06}.

\begin{figure}
\begin{center}
\includegraphics[scale=0.45,  angle=-90]{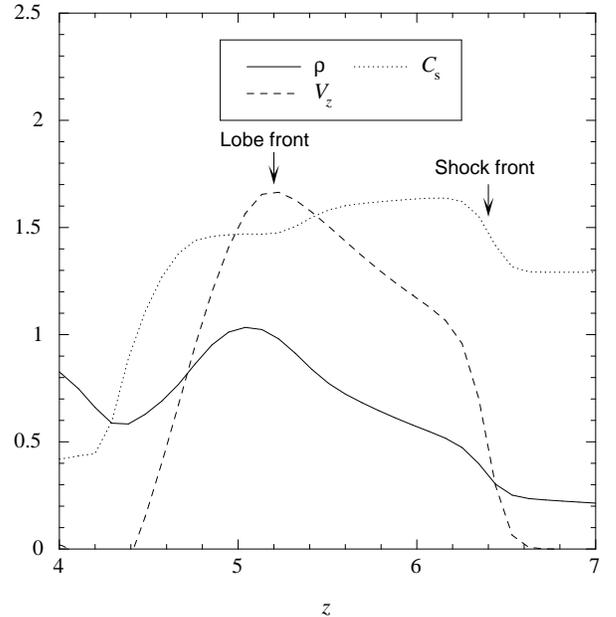} 
\caption{\label{fig:f2} Axial profiles of physical quantities along the
 $z$-axis at $t=3.0$ ($t=72$ Myrs). Density $\rho$, sound speed $C_{\rm
 s}$, and the axial velocity component $V_{z}$. The positions of the
 expanding shock and lobe fronts are shown.}
\end{center}
\end{figure}

Figure \ref{fig:f2}  exhibits physical quantities along  the $z$-axis at
$t=3.0$  ($t=72$  Myrs). The  magnetic  tower-driven hydrodynamic  shock
front can be seen around $z \sim 6.4$ in the profiles of $\rho$, $C_{\rm
s}$, and  $V_{z}$. The sound  speed jumps by  27 \% [the  temperature $T
(\propto C_{\rm s}^2)$ increases by 61 \%] in the postshocked region due
to the shock compression. The  magnetic tower front, which is identified
as the lobe front, is located at $z \sim 5.2$. There is another MHD wave
front at $z \sim 4.5$ where an increase in $C_{\rm s}$ (density/pressure
also) is  accompanied by a increase  in magnetic pressure  (not shown in
Fig. \ref{fig:f2}), indicating the reverse MHD slow mode. In later time,
this  can  steepen  into a  MHD  slow  shock  that cause  a  compression
(heating) at  the head  of magnetic tower  as seen in  Fig. \ref{fig:f1}
(see also Fig. 2 and \S 3.1 of paper II for details).

\begin{figure}
\begin{center}
\includegraphics[scale=0.45,  angle=-90]{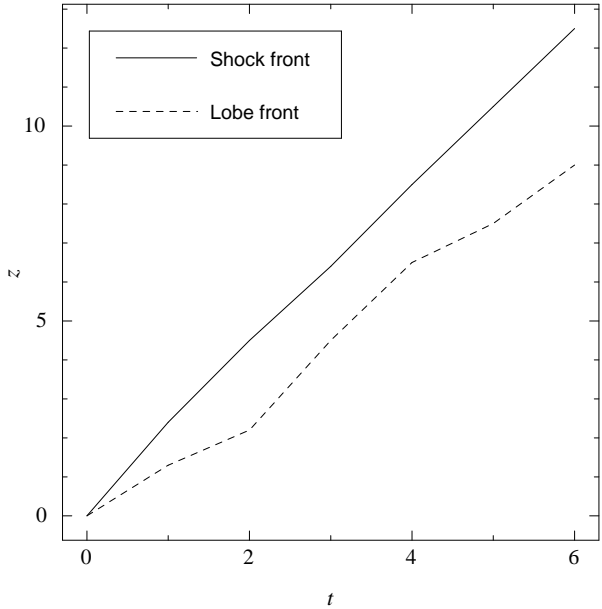} 
\caption{\label{fig:f3}  The temporal  evolution of  the shock  and lobe
front  positions along  the $z$-axis.  During the  time  evolution, both
fronts propagate  with almost constant  speeds. The shock  front expands
supersonically with  Mach number $\sim 1.63$ (relative  to the unshocked
ICM), while the lobe  front proceeds subsonically with Mach number
$\sim 0.9$ (relative to the shocked ICM cocoon).}
\end{center}
\end{figure}

Figure \ref{fig:f3} displays the positions  of the shock and lobe fronts
as  a function  of  time  along the  $z$-axis  (propagating towards  the
positive direction).   The shock front has a  constant propagation speed
$V_{\rm SH} \sim  2.1$ ($2.52 \times 10^8$ cm  s$^{-1}$) that implies a
Mach number $\sim 1.63$.   On the other hand, the magnetically-dominated
lobe   front  expands  subsonically.   This  lobe   front  also   has  a
quasi-constant  propagation speed  $V_{\rm LB}  \sim 1.5$  ($1.8 \times
10^8$ cm s$^{-1}$) that implies  a Mach number $\sim 0.91$ ($C_{\rm s}$
increases up to $\sim$ 1.64 in shocked ICM; see also Fig. \ref{fig:f2}).

\begin{figure} 
\begin{center}
\includegraphics[scale=0.65,                      angle=-90]{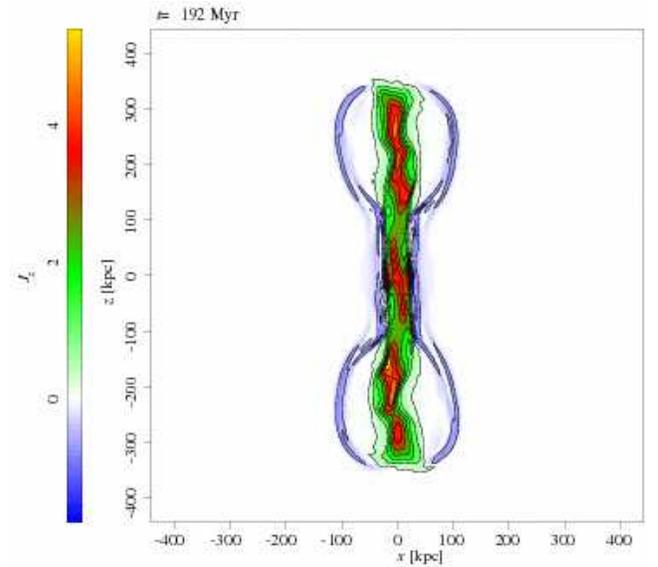}
\caption{\label{fig:f4}  Distribution  of   the  axial  current  density
$J_{z}$ in the $x-z$ plane at the same time as Fig. \ref{fig:f1}.  Two
types of growing current-driven (external/internal) modes are visible in
the lobe and jet.  }
\end{center}
\end{figure}

The current-carrying magnetic tower  jet, which possesses a highly-wound
helical  field,  is subject  to  the  current-driven instability  (CDI).
Although the  destabilization criteria will  be modified by  the ambient
gas and the RFP configuration of the tower, we find that the propagating
magnetic tower jets can develop the non-axisymmetric CDI modes.  As seen
in Fig. \ref{fig:f4}, both the internal elliptical ($m=2$) mode like the
``double helix'' and the external  kink ($m=1$) mode grow to produce the
wiggles at different locations (see also paper III for details).  We see
no disruption  owing to shear-driven, Kelvin-Helmholtz  modes and/or the
pressure-driven interchange modes.

\begin{figure} 
\begin{center}
\includegraphics[scale=0.4,                        angle=0]{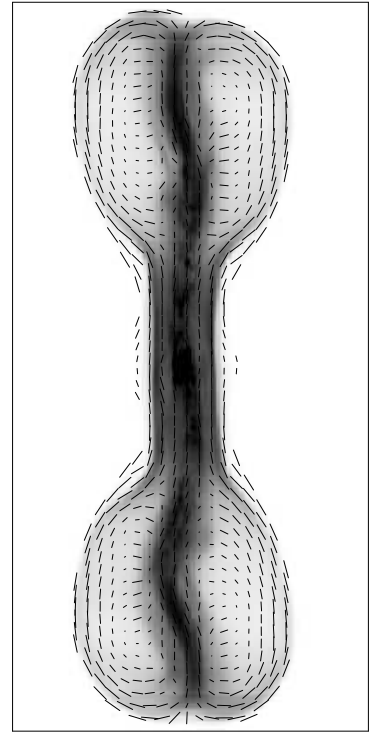}
\caption{\label{fig:f5}   Synthetic   synchrotron   surface   brightness
computed at 5 GHz (greyscale), overlaid with polarization magnetic field
vectors.   The field  vectors are  corrected for  Faraday  rotation, and
represent the field orientations  at the source.  The polarization field
is nicely aligned with the jets and lobe edges.}
\end{center}
\end{figure}

Figure \ref{fig:f5} shows the synthetic radio synchrotron intensity at 5
GHz, along  with the polarization  magnetic field vectors.   The wiggled
structures can be  observed in both the jet  and lobe. Also, filamentary
structures appear  beyond $R_{\rm c}$.  The projected  magnetic field is
aligned with the jet and along  the lobe edge.  The magnetic tower model
produces  polarization features that  are qualitatively  consistent with
those  seen  in the  Herc  A  observations  of GL03.   Furthermore,  our
dynamical   solution  of   magnetic   tower's  evolution   may  give   a
self-consistent  explanation   to  a  feasible   magnetic  configuration
discussed  in non-dynamical  models having  ``spine'' ($B_{\phi}  > B_z$
inside the lobe) -- ``Sheath'' ($B_z  > B_{\phi}$) structure in FR I jet
over 10 kpc scale as discussed in \citet[]{L81, LCB06}.

The image  contrast in our synthetic  synchrotron surface-brightness map
is approximately  2000:1 between the  brightest pixels (beams)  near the
central core  and the  faintest regions inside  the lobes.   Outside the
small bright core region, the image contrast is roughly 200:1.

When the  synthetic 5 GHz synchrotron  map is scaled to  an angular size
comparable  to  that  of Herc  A,  the  brightest  pixel (beam)  in  the
synthetic 5 GHz  surface brightness map is 875  mJy/beam.  However, away
from the small, bright region  near the core, the jets generally exhibit
flux  densities on  the  order  of 30-70  mJy/beam,  the central  sheath
exhibits flux  densities in  the range 20-30  mJy/beam, and  the darkest
parts of the lobes exhibit $\lesssim$ 1.0 mJy/beam.

Our simulated source exhibits  a fractional linear polarization, $m$, at
5  GHz ranging  from  a few  percent  in the  most magnetically  tangled
portions of the  lobes to $m \approx 30-50\%$ on  the lobe edges.  These
values are  consistent with those  typically observed in  radio galaxies
\citep{Bridle84,Muxlow91}  as well as  with the  $m$ values  reported in
GL03 for Herc A at 4848 MHz.  Fractional polarization values exceed 60\%
(roughly twice that reported in  GL03) in small, isolated pockets in the
sheath  around the  central jet  and near  the central  core,  which may
indicate  insufficient  computational   resolution  near  the  core  for
properly  capturing  cellular  depolarization.  In regions  outside  the
magnetized lobes,  where the ambient medium in  our calculational domain
has been mildly perturbed by expansion of the magnetic tower, the degree
of  polarization approaches  72\%.  This  is the  expected result  for a
region  with   a  uniform  magnetic   field  and  a   uniform  power-law
distribution of nonthermal electrons with $q = 4.5$ \citep{Rybicki79}.

\section{DISCUSSION}
\subsection{Shock and Lobe Expansions}
Some features about the shock  and lobe expansions in the simulation are
discussed by comparing with observations. A {\em Chandra} X-ray image of
the  Herc  A cluster  shows  that  it has  cavities  and  a shock  front
associated  with  the powerful  radio  lobes \citep[]{N05}.   Unusually,
these cavities show no clear connection to the radio sources, while many
cavities  are  associated with  radio  sources  in  other cluster  cases
\citep[e.g.,][]{Mc00},  indicating  that they  might  be ghost  cavities
\citep[e.g.,][]{Mc01}.

\citet{N05}'s observation reveals that  a shock front surrounds the Herc
A radio source. It is elongated  in the direction of the radio lobes and
appears  to be  a cocoon  shock  \citep[]{S74}.  Their  fitting of  a
simple hydrodynamic model to the surface brightness profile gives a Mach
number for the  shock front of $\sim 1.65$ and its  total energy $\sim 3
\times  10^{61}$  ergs  (its  mean  mechanical power  $\sim  1.6  \times
10^{46}$   ergs  s$^{-1}$)  is   estimated.   As   we  examined   in  \S
\ref{sec:RES}, the magnetic tower  drives a hydrodynamics shock front in
the ambient ICM.  Derived Mach number $M \sim 1.63$ matches quite well to
their  observation  and also  our  energy  inputs  are in  qualitatively
reasonable range.

As the radio lobes of Herc A lack anything resembling hotspots, it seems
likely that  they are expanding  at around the  sound speed, and  so are
confined  by the thermal  pressure of  the ambient  ICM, rather  than by
strong shocks \citep[]{GL04}.  To our knowledge, there may  be no direct
result revealing the lobe expansion speed  of Herc A radio lobes, but we
discuss general properties  of expanding lobes in radio  galaxy based on
\citet[]{K06}'s  argument.  It  is  believed that  the  radio lobes  are
greatly overpressurized relative to  the ambient medium, indicating that
sharp  X-ray surface brightness  discontinuities or  large jumps  in gas
temperature at the lobe edges.  A  high Mach number ($M \sim 8.5$) shock
is detected around the SW radio lobe of Centaurus A \citep[]{K03}.

However, there  is no evidence  especially for FR  I or low-power  FR II
radio   galaxies,   including  Herc   A   \citep[]{GL04}   and  3C   388
\citep[]{K06}.  If thermal conduction in  the ICM is efficient (an order
of the  Spitzer value),  then shocks may  be nearly isothermal  and thus
difficult to detect  \citep[]{F06}.  However, {\em Chandra} observations
suggest that the thermal conduction (and viscosity) of the ICM is orders
of magnitude  below the  Spitzer value \citep[e.g.,][]{V01,  K03}. Thus,
some  lobes  including  Herc   A  may  be  inflating  transonically  and
subsonically (i.e.,  $M \le 1$).   Based on hydrodynamic arguments  of a
buoyant bubble, $M \sim 0.6$ in M87 \citep[]{C02} and $M \sim 0.5 - 0.9$
in 3C  388 \citep[]{K06} are estimated.  Our  numerical result indicates
that the  magnetically-dominated lobe expands subsonically  with $M \sim
0.91$,  discussed  in  \S  \ref{sec:RES}.   This  will  also  match  the
observationally preferable picture.

\subsection{Lobe Size}
We next discuss the radial size  of the lobes $r_{\rm lobe}$ as based on
our model \citep[]{N06}.  The  poloidal magnetic fluxes and the poloidal
current $I_{z}$ in  the lobes remain well collimated  around the central
axis even  after the  system has evolved  fully.  This implies  that the
toroidal magnetic fields in the lobe region are distributed roughly as
\begin{eqnarray}
\label{eq:Biot-Savart}
B_\phi \sim I_z/r.   
\end{eqnarray}
When  the lobe experiences  sufficient expansion,  we can  expect $|B_z|
\sim   |B_{\phi}|$  at  the   outer  edge   of  the   RFP  configuration
\citep[]{F82}.  The magnetic  and background  pressures tend  to balance
each other at the tower edge. Thus, we can expect that
\begin{equation}
\label{eq:balance_edge}
\frac{B_{\phi}^{2}+B_{z}^2}{2} \sim B_{\phi}^2 \sim p_{\rm e}~,
\end{equation} 
where $p_{\rm  e}$ is the  external gas pressure  at the tower  edge. By
combining  eqs. (\ref{eq:Biot-Savart})  and  (\ref{eq:balance_edge}), we
have
\begin{equation}
\label{eq:balance_edge2}
B_{\phi}^{2} \sim \left(\frac{I_z}{r_{\rm lobe}}\right)^2 \sim p_{\rm e}~, 
\end{equation} 
which gives 
\begin{equation}
\label{eq:balance_edge3}
r_{\rm lobe} \sim I_z~p_e^{-1/2}~.  
\end{equation}

\begin{figure}
\begin{center}
\includegraphics[scale=0.45,                      angle=-90]{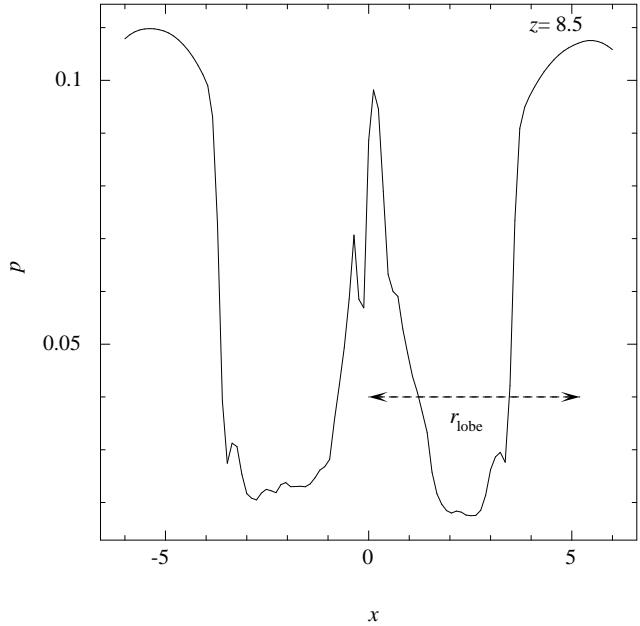}
\caption{\label{fig:f6} Transverse  profile in the  $x$-direction of the
gas pressure $p$  on $z=8.5$ (255 kpc).  The  lobe radius $r_{\rm lobe}$
calculated  via  eq.    (\ref{eq:balance_edge3})  is  also  shown.   The
pressure cavity radius and  $r_{\rm lobe}$ qualitatively agree with each
other.}
\end{center}
\end{figure}

Figure  \ref{fig:f6} shows the  distribution of  $p$ along  a transverse
line at $z=8.5$ (255 kpc), when the lobe has a maximum radius (after 192
Myr, as  in Figs.  \ref{fig:f1} and  \ref{fig:f4}). We can  see that the
external gas  pressure begins  to decrease around  $x \sim 5$  (150 kpc)
toward the central  ($z$) axis.  The pressure cavity  corresponds to the
lobe  of our magnetic  tower. We  can estimate  the lobe  radius $r_{\rm
lobe}$   predicted  by   eq.   (\ref{eq:balance_edge3})   from  physical
quantities in simulation.  From Fig.  \ref{fig:f2}, we take $J_z \sim 3$
($4.5  \times  10^{-24}$ A/cm$^{-2}$)  within  a  radius  $r \sim  0.75$
($22.5$ kpc), which makes $I_{z} \sim 1.68$ ( $5 \times 10^{17}$ A).  As
suggested by Fig.  \ref{fig:f4}, we estimate $p_{\rm e} \sim 0.11$ ($2.5
\times  10^{-11}$ dyn  cm$^{-2}$)  for the  external  gas pressure.   We
therefore obtain $r_{\rm  lobe} \sim 5.2$ or roughly  156 kpc.  Thus, we
find that  the depression radius  in Fig. \ref{fig:f6}  is qualitatively
consistent with $r_{\rm lobe}$ and is comparable with the physical scale
seen in radio observations \citep{GL03}. Based on this numerical result,
we estimate that one simulated lobe of Herc-A may contain a total energy
$\sim 1.8 \times  10^{61}$ ergs, out of which  $\sim 1.0 \times 10^{61}$
ergs  are in  magnetic energy,  $\sim 1.4  \times 10^{60}$  ergs  are in
kinetic  energy, and  $\sim  6.6  \times 10^{60}$  ergs  are in  thermal
energy,  respectively.  In  order to  explain  the lobe  radius by  this
magnetically  dominated lobe  system, an  axial current  $\sim  5 \times
10^{17}$ A, which is flowing along the jet central axis, is needed.

\section{CONCLUSIONS}
By  performing 3D MHD  simulations, we  have investigated  the nonlinear
dynamics of  magnetic tower  jets in a  galaxy cluster  environment.  In
this fourth paper of the  series, we have applied our numerical modeling
of the magnetic tower to a specific source, Hercules A.

To our knowledge,  this is the first trial of  discussing Hercules A jet
by  using   a  magnetohydrodynamic  model.    Projected  magnetic  field
distributions associated with the jets and lobes of Hercules A have been
revealed  in  recent  radio  observation  and  therefore,  the  dynamics
including magnetic fields plays a role in understanding this source.  In
the present  paper, we have investigated the  jet-driven shock, jet/lobe
transition,  wiggling,  and  projected  magnetic field  distribution  of
Hercules  A and, in  general, these  feasutres can  be explained  by our
magnetic tower model.

Our conclusions are summarized as follows:
\begin{enumerate}
\item A transition from the narrow jet  to the fat lobe can occur at the
cluster core radius.  This interesting  feature can also be confirmed in
recent observations  \citep[]{GL03, GL04}.   In our models,  the tightly
collimated helical  field configuration of a magnetic  tower jet expands
abruptly  beyond the  core radius  due  to the  decreasing external  gas
pressure.  Thus,  the pressure profile  of the surrounding ICM  plays an
important role in determining the morphology of a magnetic tower jet.

\item  The   expanding  magnetic   tower  jet  produces   the  preceding
hydrodynamic shock wave  in the ambient ICM. This  can be interpreted as
AGN-driven shock in X-ray observation  and the derived shock Mach number
is identical to \citet[]{N05}'s  observation result.  The magnetic tower
lobes expand subsonically and thus no  hot spots are produced at the end
parts of lobes. This may be a general understanding of FR I or low power
FR II radio galaxies \citep[]{K06}.

\item The  size of the magnetic  tower lobe can  be accurately estimated
from the  jet axial current  and the external  gas pressure at  the lobe
edge. The  estimated lobe  size derived by  our numerical  simulation is
qualitatively   comparable    with   the   physical    scale   seen   in
\citet[]{GL03}. In  our model, if the  background gas has  a steep slope
($\beta$) in the King profile,  the magnetic jet can potentially lead to
a huge magnetic tower lobe.

\item Magnetic tower jets, which have a reversed magnetic pinch profile,
are subject to current-driven instabilities.  Specifically, the apparent
distortions by  the non-axisymmetric  kink mode are  visible at  the jet
body inside  the lobe rather than  the central core.  By expanding
the lobe the  edge of axial currents could be  free against the exciting
external kink mode; this situation  may be suppressed inside the cluster
core radius.

\item The synthetic polarimetry of our magnetic-tower jets is consistent
with the gross polarization features observed in the jets of Hercules A.
The magnetic tower model produces projected $\bolB$-vector distributions
that are similar to those  observed in the global extragalactic jet-lobe
system, especially along the jets and the lobe edges.

\item Magnetic tower  jets may produce a self-consistent  picture of the
field  configuration discussed in  several non-dynamical  model of  FR I
jets.  It consists  of toroidally  dominated spine  inside the  lobe and
poloidally dominated sheath at the lobe edge (a closed poloidal magnetic
flux with a  toroidal magnetic component, rather than  one side directed
helical configuration).
\end{enumerate}

\acknowledgments  Helpful discussions with  Philipp Kronberg  and Steven
Diehl  are gratefully  acknowledged.   The authors  thank the  anonymous
referee for  helpful suggestions.  This work was  carried out  under the
auspices  of the National  Nuclear Security  Administration of  the U.S.
Department of  Energy at Los  Alamos National Laboratory  under Contract
No.   DE-AC52-06NA25396.  It  was supported  by the  Laboratory Directed
Research and Development Program at LANL and by IGPP at LANL.



\end{document}